\documentclass[letter]{aa} 

\usepackage{graphicx}
\usepackage{txfonts}
\usepackage{graphicx}
\usepackage{color}
\usepackage{float}

\usepackage{graphicx}
\usepackage{txfonts}
\usepackage{hyperref}
\usepackage[normalem]{ulem}
\hypersetup{colorlinks = true, citecolor = {blue}, urlcolor= {blue}}

\usepackage{natbib}
\bibpunct{(}{)}{;}{a}{}{,} 
\usepackage{txfonts}
\usepackage{amssymb}
\usepackage{amsfonts}
\usepackage{tabularx}
\usepackage{graphicx}
\usepackage{ulem}
\usepackage{array}
\usepackage{color}
\usepackage{booktabs}
\usepackage[fleqn]{amsmath}
\usepackage{subfigure}
\usepackage{diagbox}
\usepackage{siunitx}

\begin{document}

   \title{Mitigating potentially hazardous asteroid impacts revisited}

   \subtitle{}

   \author{Zs. Regály
          \inst{1,2}
          \and
          V. Fröhlich
          \inst{1,2,3}
          \and
          P. Berczik\inst{1,2,4,5}
          }

\institute
{
{Konkoly Observatory, Research Centre for Astronomy and Earth Science, Konkoly-Thege Mikl\'os út 15-17, 1121, Budapest, Hungary\\
\email{\href{mailto:regaly@konkoly.hu}{regaly@konkoly.hu}} }
\and
CSFK, MTA Centre of Excellence, Budapest, Konkoly Thege Miklós út 15-17., H-1121, Budapest, Hungary
\and
{E\"otv\"os Lor\'and University, P\'azm\'any P\'eter s\'et\'any 1/A, 1117 Budapest, Hungary}\\
\email{\href{mailto:frohlich.viktoria@csfk.org}{frohlich.viktoria@csfk.org}}
\and
{Zentrum f\"ur Astronomie der Universit\"at Heidelberg, Astronomisches Rechen-Institut, M\"{o}nchhofstr. 12-14, 69120, Heidelberg, Germany}
\and
{Main Astronomical Observatory, National Academy of Sciences of Ukraine, 27 Akademika Zabolotnoho St, 03143 Kyiv, Ukraine \email{\href{mailto:berczik@mao.kiev.ua}{berczik@mao.kiev.ua}}
}
}

\date{Received \today; accepted }


 
  \abstract
   {Potentially hazardous asteroids (PHA) in Earth-crossing orbits pose a constant threat to life on Earth. Several mitigation methods have been proposed, and the most feasible technique appears to be the disintegration of the impactor and the generation of a fragment cloud by explosive penetrators at interception. However, mitigation analyses tend to neglect the effect of orbital dynamics on the trajectory of fragments.}
   {We aim to study the effect of orbital dynamics of the impactor's cloud on the number of fragments that hit the Earth, assuming different interception dates. We investigate the effect of  self-gravitational cohesion and the axial rotation of the impactor.}
   {We computed the orbits of $10^5$ fragments with a high-precision direct N-body integrator of the eighth order, running on GPUs. We considered orbital perturbations from all large bodies in the Solar System and the self-gravity of the cloud fragments.}
   {Using a series of numerical experiments, we show that orbital shear causes the fragment cloud to adopt the shape of a triaxial ellipsoid.
   The shape and alignment of the triaxial ellipsoid are strongly modulated by the cloud's orbital trajectory and, hence, the impact cross-section of the cloud with respect to the Earth.
   Therefore, the number of fragments hitting the Earth is strongly influenced by the orbit of the impactor and the time of interception.
   A minimum number of impacts occur for a well-defined orientation of the impactor rotational axis, depending on the date of interception.}
   {To minimise the lethal consequences of an PHA's impact, a well-constrained interception timing is necessary. A too-early interception may not be ideal for PHAs in the Apollo or Aten groups. Thus, we find that the best time to intercept PHA is when it is at the pericentre of its orbit.}

   \keywords{Celestial mechanics --
   Earth --
                Meteorites, meteors, meteoroids --
                Methods: numerical
               }

   \maketitle
%

\section{Introduction}

The Cretaceous-Paleogene extinction event wiped out about 75 per cent of all species \citep{Raupetal1982}: all non-avian dinosaurs died out, while all ammonites, plesiosaurs, and mosasaurs disappeared from the seas \citep{Fastovskyatal2005}. 
This left mammals and birds to dominate the land.
\cite{Kelly1953} proposed and later \citet{Alvarez1980} and \citet{Smit1980} confirmed the hypothesis that these events were caused by the impact of a large asteroid of 10-15 km diameter on the Earth 66 million years ago \citep{Renneetal2013}.

Near-Earth objects (NEOs) are asteroids or comets with a perihelion distance, $q$, less than 1.3~au. 
Near-Earth asteroids (NEAs), are divided into four groups\footnote{See details at NASA's Near Earth Objects Basics page: https://cneos.jpl.nasa.gov/about/neo\_groups.html}
(Atira, Aten, Apollo, and Amor) based on their semi-major axis and eccentricity.
Asteroids with a diameter of more than 140~m in the Apollo and Aten groups  having Earth-crossing orbits, are classified as potentially hazardous asteroids (PHAs). 
We note that comets crossing the Earth's orbit can also be considered PHAs.

Although the recurrence interval of an impact event with a diameter of 10-15\ km is about 0.2 billion years, a smaller impact event with a diameter of 1\ km can be expected to occur every 0.6 million years \citep{Collinsetal2005}.

In a recent review by \citet{Lubinetal2023}, six methods of mitigating PHA impact  are discussed.
The most promising technique is the use of an array of penetrators combined with nuclear explosives that disintegrate the asteroid impactor at the so-called interception event into a gravitationally unbound cloud of fragments with sufficient energy to spread, (see, e.g., \citealp{Kaplingeretal2010,Sanchezetal2008,Kingetal2021}). 
The shock waves caused by the impact of fragments of an asteroid with a diameter of 1\,km can be decorrelated (the impact events are dispersed in time) in the Earth's atmosphere to a sufficient degree to reduce the threat.
Furthermore, assuming an intercept time of more than about 2.5 months, a fragment expansion speed of about $1\,\mathrm{m\,s^{-1}}$ is sufficient to miss Earth.
However, for a 10\,km diameter asteroid, the amount of energy released by the shock wave and even the dust production becomes large enough to overwhelm our atmospheric shield.
Therefore, a much earlier intercept will be required to make the cloud of fragments grow large enough, so that only a small fraction of them end up hitting the Earth.

To predict the number of fragments that hit Earth, \citet{Lubinetal2023} use an analytic approximation neglecting the effect of orbital dynamics on the trajectory of fragments.
In this letter, we show that the fragment cloud is strongly distorted by the orbital shear, so that classical analytical approximations cannot be used to estimate the size of the impact.
Therefore, in agreement with the results of \citet{Kingetal2021}, using a series of high-precision N-body simulations is necessary.

\section{Numerical simulations}

A fully gravitationally interacting many-body system is used to model the impact event. 
The impactor is resolved as $10^5$ spherical particles of equal mass and size. 
The individual orbits of the impactor fragments, perturbed by the Sun and the major bodies of the Solar System (planets, Moon, and Pluto), are computed with a high-precision direct N-body integrator.
We performed two sets of simulations: 1) a gravitationally non-interacting impact cloud to reveal the effect of orbital shear and 2) a fully interacting impact cloud, for which the gravity of the fragments is also considered. 

We used our GPU–aided code HIPERION\footnote{High Precision Integrator for N-body, HIPERION code can be accessed at \url{https://konkoly.hu/staff/regaly/research/hiperion.html}}, utilising an eighth-order Hermite scheme described in \citet{NitadoriMakino2008}.
Based on the predictor-corrector relative acceleration of bodies, a shared adaptive timestep method is used (see details in Appendix~\ref{sec:apx-integrator}). 
Every time step is controlled such that simulations provide better positional accuracy than the diameter of the impactor.

\subsection{Ephemerides and impact events}

To take into account the perturbation effects of the Solar System bodies on the impactor orbit, we obtained the position and velocity vectors of the planets, Pluto, and the Moon from the NASA JPL Horizons service\footnote{ \url{https://ssd.jpl.nasa.gov/horizons/app.html\#/}} at the time of a hypothetical impact event.
Throughout the investigation, 12:00am 24/06/2030 (JD~2462677.0) was used as the date of the impact event.
The Astroquery Python package was used to retrieve the ephemerides \citep{ginsburg-etal-2019}.

We considered an asteroid and a comet-type impactor, which have the same diameter but different internal densities and impact velocities (physical properties are shown in Table~\ref{tab:param}).
We note that the diameter of the impactors is chosen to ensure their survival even in sungrazing orbits \citep{Sekanina2003}.
Central collisions between Earth and the impactors are assumed, meaning that the impactor's velocity vector points towards the Earth's center of mass.

The properties of the impact are defined by three parameters: $v_\mathrm{imp}$ as the impact velocity; $\delta$ as the angle between the velocity vector of the Earth and that of the impactor in the plane of the Earth's orbit; and 
$\phi$ as the angle between the velocity vector of the impactor and the orbital plane of the Earth.
$\delta=0^\circ$ or $\delta=180^\circ$ correspond to a head-on or a rear-end collision with Earth, respectively.
The calculated orbital elements of the impactor models are presented in Appendix~\ref{sec:apx-orbits}. 

\begin{table}
        \centering
        \caption{Parameters of the asteroid and comet type impactors.}
        \label{tab:param}
        \begin{tabular}{lcc} 
                Physical property                    & Asteroid             & Comet \\
  \hline
        Diameter (m)            & 1000 & 1000  \\ 
        Density $(\mathrm{kg\,m^{-3}})$       & $2.6\times10^3$       & $0.65\times10^3$   \\
        Mass (kg)               & $1.36\times 10^{12}$  & $3.40\times10^{11}$ \\

        $v_\mathrm{imp}\,(\mathrm{km\,s^{-1}})$ & 20                    & 40                 \\
        $v_\mathrm{esc}\,(\mathrm{m\,s^{-1}})$ & 0.625 & 0.301\\
        $P_\mathrm{min}$ (hour) & 2.04657 & 4.09315 \\
        $v_\mathrm{surf}\,\mathrm{(m\,s^{-1})}$ & 0.425687 &  0.212844 \\
                \hline
  \end{tabular}
\end{table}

\begin{figure*}
    \centering
    \includegraphics[width=1\columnwidth]{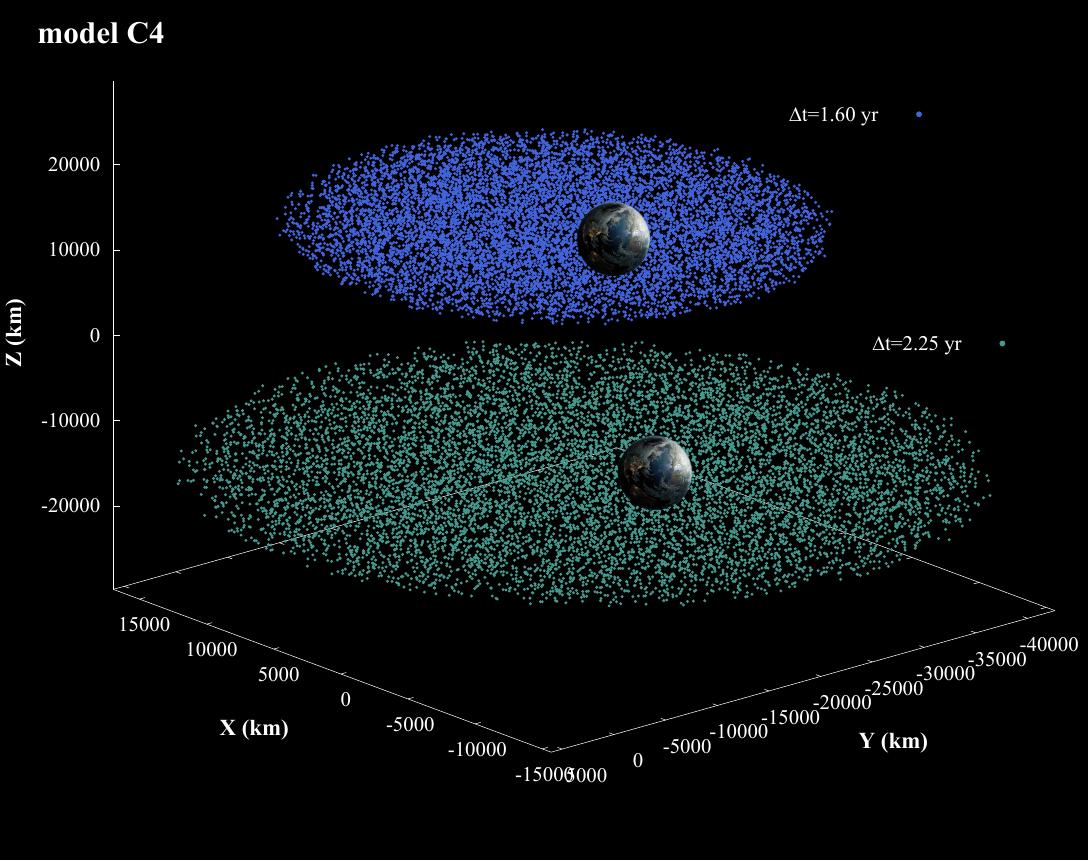}
    \includegraphics[width=1\columnwidth]{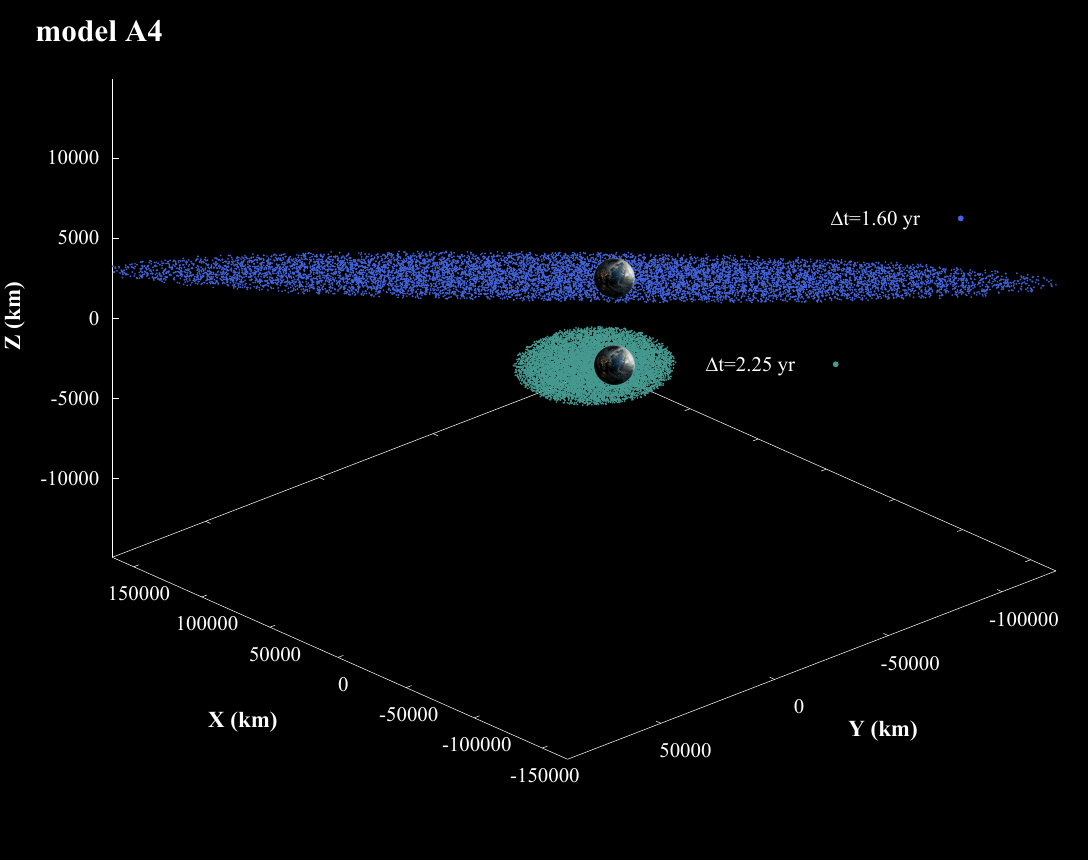}
    \caption{Fragment clouds immediately before the impact to the impact for models C4 (left) and A4 (right).
    Two interception dates are shown: $\Delta t=1.6$\,yr and $2.2$\,yr.
    Note: the Z axis scale is independent of the X and Y axis scales and the fragment clouds are shifted arbitrarily for better visualisation.
    }
    \label{fig:cross}
\end{figure*}

\subsection{Backward-forward integration, fragment cloud, and rotation} 

For the modelling of an impact event, a backward integration in time is first performed by reversing the velocity of the impactor as well as all other Solar System bodies at the impact event. 
The difference between the interception date ($t_\mathrm{int}$) and the impact event time ($t_\mathrm{imp}$) defines the length of the backward integration. 
We examine a series of $\Delta t=t_\mathrm{imp}-t_\mathrm{int}$ sampled over the interval $1/48-3$~years by 144 different integration times.
At the end of the backward integration, that is, at the interception, the velocity vectors of all celestial bodies are reversed again.
Note that the backward-forward integration method is similar to the one used in \citet{Kingetal2021}.

Using the particle packing algorithm presented in \citet{BaranauTallarek2017}, the impactor is disintegrated into $1.04739\times10^5$ individual monodisperse particles assuming a uniform size distribution for simplicity) at the point of impact.
The diameter of the individual fragments is about $18$\,m and the corresponding porosity is measured to be $0.39$. 

Fragments inherit the impactor velocity components at interception position.
To avoid the collapse of the fragment cloud by its own gravity, an additional spherically symmetric velocity distribution proportional to the local escape velocity is superimposed.
In the applied expansion model, the velocity vectors of the fragments are given by the equation: 
\begin{equation}
{\bf v_\mathrm{exp}}(R_\mathrm{i})=\left(\frac{{\bf R}_\mathrm{i}}{R_\mathrm{i}}\right)v_0 v_\mathrm{esc}(R_\mathrm{i}),
\label{eq:vexp}
\end{equation}
where ${\bf R}_\mathrm{i}$ and $R_\mathrm{i}$ are the position vector and distance of the i-th fragment measured from the centre of mass of the impactor and $v_\mathrm{esc}(R_\mathrm{i})$ is the local escape velocity at $R_\mathrm{i}$.
Four expansion models with different intensities are investigated assuming $v_0=1,\,2,\,5,$ and $10$.

In order to model the effect of the axial rotation of the impactor, simulations are performed assuming a minimum rotational period, which is determined by the maximum rotational speed of an impactor made of debris and held in place by gravity alone.
This period is calculated based on \citet{PravecHarris2000} as $P_\mathrm{min}=3.3 / \sqrt{\rho}$, where $\rho$ is the internal density of the impactor.
The corresponding rotational velocity at the surface of the impactor  with a diameter of $D$ is $v_\mathrm{surf}=(D\pi)/{3.3}\sqrt{\rho}$.

We assume that each fragment receives a tangential kick due to axial rotation at the moment of expansion onset. 
The kick velocity, 
\begin{equation}
    {\bf v}_\mathrm{kick}(R_\mathrm{i})= v_\mathrm{surf}  \frac{2R_\mathrm{i}}{D} {\bf M}(\alpha,\beta) \left(\frac{{\bf R}_\mathrm{i,\perp}}{R_\mathrm{i}}\right) ,
    \label{eq:rotvel}
\end{equation}
is added to the velocity of the fragment Eq.~\ref{eq:vexp}, where ${\bf R}_\mathrm{i,\perp}$ is the tangential vector at a position, $R_\mathrm{i}$, and ${\bf M}(\alpha,\beta)$ is the Euclidean rotation matrix.
The orientation of the impactor's rotation axis is defined by two angles: $\alpha$ measures the inclination from the normal to the impactor's orbital plane, ${\bf n}_\mathrm{o}$, whereas $\beta$ measures the angle of rotation along ${\bf n}_\mathrm{o}$.
The effect of the rotation is studied with $50\times50$ different $\alpha$ and $\beta$ values in the range $[0,\pi]$.

The shape of the fragment cloud is determined as it approaches Earth. 
This is done by fitting a triaxial ellipsoid to the particle distribution.
The principal axis transformation of the moment of inertia tensor (see details in Appendix~\ref{apx:triax}) is used to obtain the best-fit ellipsoid.
The parameters to be determined are as follows: first Euler angle $\theta$, the size of the axes of the ellipsoid ($a,\,b,$ and $c$), and the oblateness $b/a$ and $c/b$ of the cloud.

\section{Results and discussions}
\label{sec:results}

\subsection{Importance of orbital dynamics}

Our simulations, in agreement with \citet{Sanchezetal2008}, show that orbital shear transforms the initially spherical fragment cloud into a triaxial ellipsoid.
The orientation of the largest semi-major axis of the best-fit ellipsoid changes during the cloud's flight and remains in its  orbital plane.
We will show that the cloud is shaped by the orbital shear (pericentre or apocentre passages), resulting in a non-trivial dependence of the impact number on the date of interception.
Figure\,\ref{fig:cross} shows the fragment clouds immediately before the impact for the models~C4 and A4, assuming two different dates of interception.
Surprisingly, the size of the late intercept cloud is larger than the early one because it's more elongated for model A4.
As a result, more fragments can hit the Earth for an early interception.

\begin{figure*}
    \centering
    \includegraphics[width=2\columnwidth]{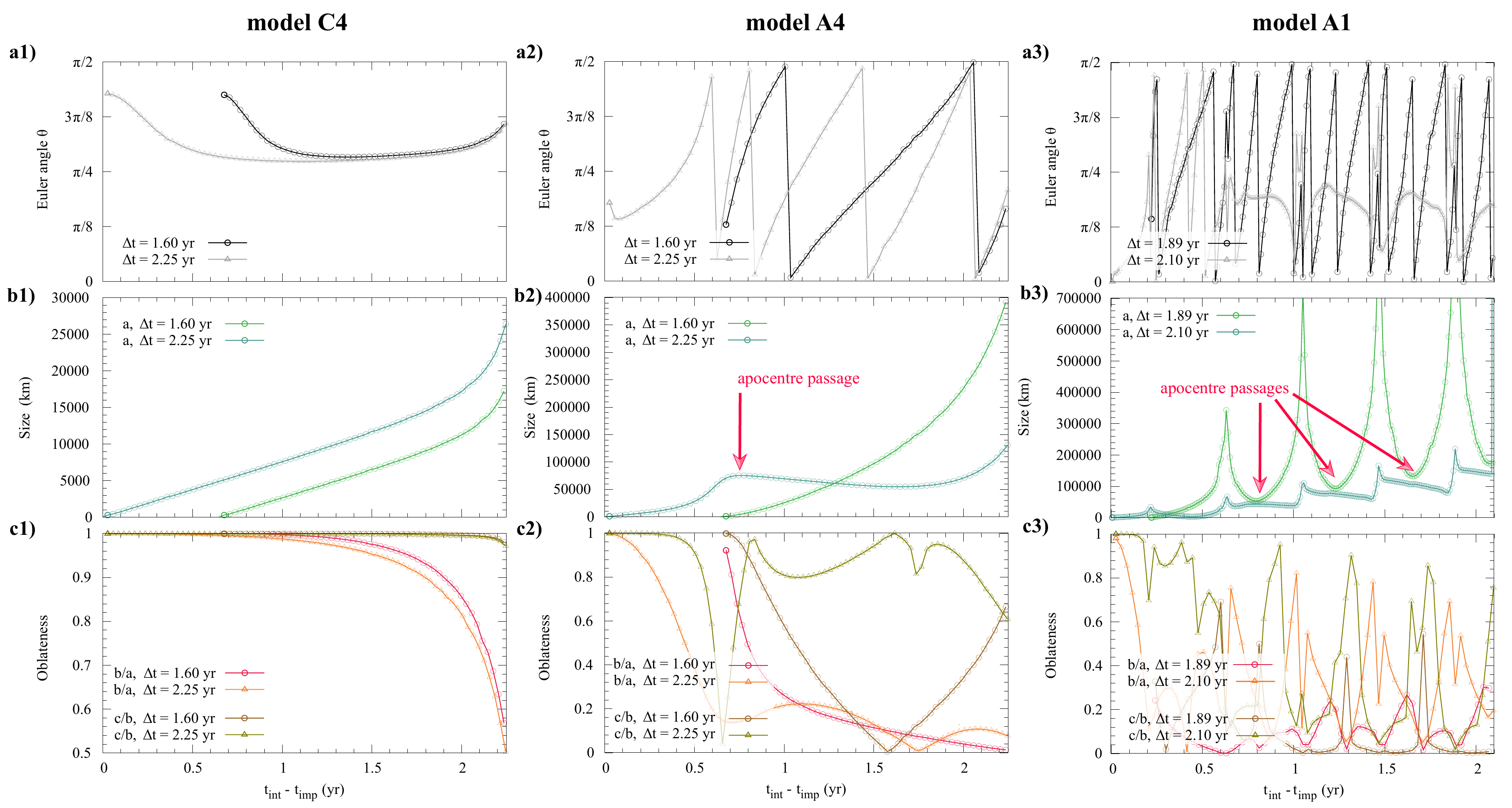}
    \caption{Variation of the parameters of the ellipsoid that best fits the cloud of fragments as it approaches the Earth. \emph{Top panels}: First Euler angle $\theta$. \emph{Middle panels}: Size of the semi-major axis, $a$. \emph{Bottom panels}: Oblateness (ratios of the semi-minor axes b/a and c/b).
    Two interception dates are compared: $\Delta t=1.6$\,yr vs. $2.25$\,yr for models C4 and A4, $\Delta t=1.885$\,yr vs. $2.10$\,yr for model A1.}
    \label{fig:triax}
\end{figure*}

First, we consider a comet-type impactor approaching Earth on a hyperbolic orbit (model C4), assuming two interception dates $\Delta t=1.6$\,yr and 2.25\,yr.
The left panels of Fig.~\ref{fig:triax} show the variations in the parameters of the best-fit ellipsoid as the impactor approaches Earth.
As the fragment cloud passes by the apocentre, it is deflected by $\sim-25^\circ$, and later of $\sim15^\circ$ during its final approach to Earth (panel~a1).
The latter deflection is caused by the Earth's orbital perturbation (see Fig.~\ref{fig:close-perturb}).
The semi-major axis of the cloud grows to $2.592\times10^4$~km and $5.281\times10^4$~km prior to the impact for the two models (panel~b1).
The fragment cloud becomes an oblate spheroid with $c/b\simeq1$ and $b/a\simeq0.5$ (panel~c1).
The semi-major axis of the ellipsoid grows monotonically, and the associated increase in volume is $V/V_0= 1.303\times10^{13}$ and $3.579\times10^{13}$. 
As a consequence an earlier interception results in a more diluted cloud (left panel in Fig.~\ref{fig:cross}). 
The measured number of fragments that impact Earth, that is, the impact numbers are $N_\mathrm{imp}=55~529$ and $30~240$ for $\Delta t=1.6$\,yr and $2.25$\,yr, respectively.

Next, we demonstrate the importance of  the effect of the orbital shear for an asteroid-type impactor in an elliptical ($e=0.669$) orbit in model A4, with $\Delta t=1.6$\,yr and 2.25\,yr interception dates (middle panels of Fig.~\ref{fig:triax}).
We note that the impactor is at pericentre and apocentre during interception.
In both models, the oblate cloud completes a full rotation during its orbit around the Sun (panel~a2).
The semi-major axis of the cloud at impact is $a=4.023\times10^5$\,km for the late and $a=1.3086\times10^5$\,km for the early interception date (panel~b2).
Although the fragment cloud expands for a longer period of time, it is almost an order of magnitude smaller before impact in the latter case (right panel in Fig.~\ref{fig:cross}).
As a result, the corresponding impact numbers are $N_\mathrm{imp}=6~951$ and $21~210$.
This can be explained by the fact that the orbital velocity increases (decreases) between  the apocentre and the pericentre (and vice versa), causing compression/decompression of the cloud (see previous findings of \citealp{Sanchezetal2008}).
During the apocentre or pericentre passages, the cloud oblateness ($b/c$) changes severely and the shortest axis ($c$) decreases by two orders of magnitude (panels~c2).

Finally, we consider a highly eccentric impact orbit (model A1) assuming $\Delta t=1.885$\,yr and $2.10$\,yr (right panels of Fig.~\ref{fig:triax}). 
The clouds are intercepted at the pericentre and the apocentre, respectively.
In these particular cases, the orbital period is short ($P=0.419$\,yr), thus several pericentre and apocentre passages occur before impact, which are associated with several full rotations of the cloud (panel~a3) and a periodic compression and decompression (panel~b3). 
Noticeably, the compression and the following decompression magnitude is more pronounced for the late interception.
The final size of fragment cloud is smaller for the earlier interception, that is, $a=3.093\times10^5$\,km and $a=2.422\times10^5$\,km for $\Delta t=1.885$\,yr and $2.1$\,yr, respectively. 
The corresponding impact numbers are $N_\mathrm{imp}=85~415$ and $N_\mathrm{imp}=103~601$.

\subsection{Interception dates and fragment's self-gravity}
\label{sec:int-date}

The number of impacts $N_\mathrm{imp}$ as a function of $t_\mathrm{int}-t_\mathrm{imp}$ assuming different expansion velocities ($1\leq v_0\leq10$) are measured for all models presented in Table~\ref{tab:orbelem} and compared with 
an analytical prediction of $N_\mathrm{imp}$ (Appendix~\ref{apx:analytic}), as shown in Fig.~\ref{fig:Nimp}.
The effect of fragment's self-gravity is also investigated.

The analytical estimation and the numerical measurement of $N_\mathrm{imp}$ are very similar for comet-type impactors approaching on a hyperbolic trajectory (models C3 and C4), assuming $v_0\geq2$ (panels~C3 and C4 of Fig.~\ref{fig:Nimp}).
We note, however, that the analytical predictions underestimate the numerical measurements by about a factor of 2 for self-gravitating models with $v_0=1$, while they slightly overestimate the numerical simulations for non-self-gravitating models.
For an asteroid-type impactor on a hyperbolic orbit (model A3) or on an elliptic orbit (models A2 and A4), the numerical measurements show a growing discrepancy with $\Delta t$ up to about a factor of 2 (panels~A2 and A3 of Fig.~\ref{fig:Nimp}).
We note that comet-type impactors in models C2-C4 do not show this phenomenon.

Models A1 and C1 demonstrate that $N_\mathrm{imp}$ cannot be predicted by a simple analytical estimation (e.g. with Eq.\,(\ref{eq:Nimp}) for NEOs in the Aten group (panels~A1 and C1 in Fig.~\ref{fig:Nimp}).  We find that
$N_\mathrm{imp}$ fluctuates as a function of $\Delta t$ with an amplitude proportional to $v_0$ and a period comparable to the impactor's orbital period.
Variations in the final size of the fragment cloud caused by intercepting at different orbital positions can explain the variation in the impact number.
This phenomenon is more robust for fast expanding clouds, namely, for $v_0>2$ models.

Generally, self-gravity of the fragments has a tendency to increase the number of impacts.
This effect is strong for models assuming $v_0=1$, while it is very weak for a relatively fast cloud expansion, $v_0>2$, (see the mismatch of the dotted lines and symbols in Fig.~\ref{fig:Nimp}).
This can be explained by the fact that the volume of the fragment cloud is larger (by a factor of 4.8 for model A4) if the fragment's self-gravity of the cloud is neglected.
The self-gravity of the fragment is much weaker for comet-type impactors because they are less massive than asteroid-type impactors.

\subsection{Orientation of axial rotation matters}
\label{sec:rot}

To investigate the effect of the impactor's axial rotation, we run additional non-self-gravitating fragment cloud models that account for an additional velocity component due to rotation based on Eq.~\ref{eq:rotvel}.
The expansion velocity is $v_0=2$, in which case the non-self-gravitating models give an accurate solution (see above).
Simulations are performed assuming $\Delta t=0.6$\,yr and $1$\,yr to study the combined effect of the interception date and axial rotation.

Figures~\ref{fig:Ast-spin} and \ref{fig:Comet-spin} show the distribution of $N_\mathrm{imp}$ in the $\alpha,\-\beta$ plane, measured for asteroid-type and comet-type impactors, respectively. 
In general, a reduction of about 30 per cent in $N_\mathrm{imp}$ is achieved by axial rotation with a minimum period.
We find that $N_\mathrm{imp}$ depends on the orientation of the rotation axis (with a few per cent variation) and there is a well-defined minimum number of impacts for each scenario. 
A comparison of the left-hand and right-hand columns shows that the effect of the date of interception is weak for models A1, A3, C1, C2, and C3.
However, for models A2, A4, and C4, we find that $N_\mathrm{imp}$ reaches its minimum value at different points, depending on the interception date.

\section{Conclusions}

Here, we present a series of numerical simulations predicting the number of fragments that hit Earth from disintegrated asteroid-type ($v_\mathrm{imp}=20\,\mathrm{km\,s^{-1}}$) and comet-type ($v_\mathrm{imp}=40\,\mathrm{km\,s^{-1}}$) impactors with a diameter of 1000\,m. 
We studied the effect of orbital shear, the interception date, and the axial rotation of the impactor.
Based on the numerical simulations carried out, we have drawn the following conclusions.

\begin{enumerate}
\item The fragment cloud is shaped by orbital shear, making an oblate spheroid (for a hyperbolic orbit) or a triaxial ellipsoid (elliptical orbit). 
The pericentre and apocentre passages modulate the size and the oblateness of the ellipsoid, as well as, the volume of the cloud, which affects the number of fragments that hit the Earth.

\item To predict the number of impacts on Earth the orbital dynamics must be correctly calculated. 
Self-gravity of the fragments is important if the cloud expands at a rate slower than twice the local escape velocity. 
In some cases ($180^\circ\leq\delta \leq270^\circ$), the analytical approximation is feasible for comet-type impactors if the cloud expands faster than the local escape velocity.

\item If the impactor is on a highly eccentric orbit, premature disassembly is undesirable. The best time to disintegrate the impactor is when it is at the pericentre of its orbit.

\item The impact number is affected by the orientation of the spin axis of the impactor prior to disassembly. For a well-defined orientation, which can depend on the date of interception, a minimum number of impacts can be found.
\end{enumerate}

Finally, since early interception may not work in all cases, NEOs in the Apollo or Aten groups with orbital periods longer than one year on highly eccentric orbits  may pose a major threat. 
To minimise the number of fragments hitting Earth, a sensitive interception timing is necessary.
The impact hazard could be mitigated by the proven existence of an optimal orientation of the impactor's rotational axis.We propose and explore a possible approach to this issue  in a forthcoming study.

\begin{acknowledgements}

VF acknowledges financial support from the ESA PRODEX contract nr. 4000132054.

The work of PB was supported by the Volkswagen Foundation under the special stipend No.~9B870. 

PB acknowledges the support within the grant No.~AP14869395 
of the Science Committee of the Ministry of Science and 
Higher Education of Kazakhstan ("Triune model of Galactic 
center dynamical evolution on cosmological time scale"). 

The work of PB was also supported under the special 
program of the NRF of Ukraine Leading and Young 
Scientists Research Support - ”Astrophysical 
Relativistic Galactic Objects (ARGO): life cycle 
of active nucleus”, No.~2020.02/0346.

\end{acknowledgements}




\bibliographystyle{aa}
\bibliography{references} 




\appendix

\section{Impactors' orbital elements} 
\label{sec:apx-orbits}

The orbital elements of the asteroid-type and comet-type impactors determined at the time of impact are given in Table~\ref{tab:orbelem}.
However, it should be noted that the impactor does not approach on these orbits, as the Earth strongly perturbs the impactor's orbit during a close encounter.

\begin{figure}[H]
    \centering
    \includegraphics[width=0.9\columnwidth]{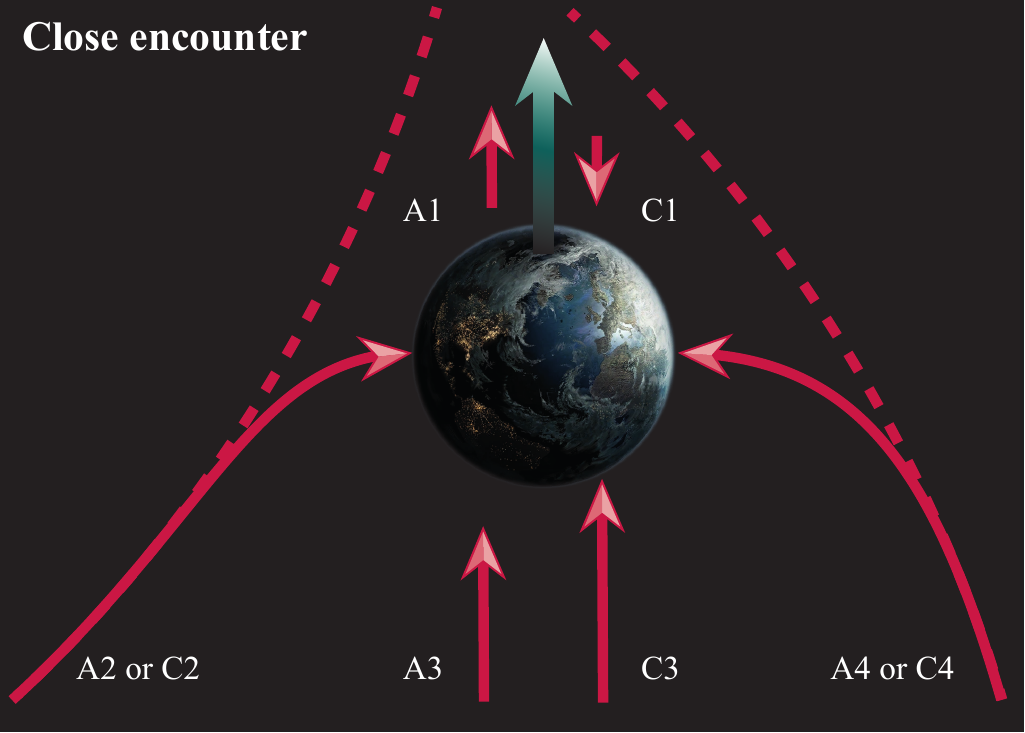}
    \caption{Perturbation of impactor orbits during a close encounter with Earth. The impactor trajectories shown as dashed red lines (models A2, C2, A4, and C4) are strongly perturbed. However, the trajectories for head-on and rear-end impact (models A1, C1, A3, and C3) are not perturbed by the Earth.}
    \label{fig:close-perturb}
\end{figure} 

In order to determine the unperturbed orbit of the impactor, we conducted a backward integration in time, for instance, to the point of interception and then re-calculated the orbital elements.
This way, the effect of the Earth's perturbation is taken into account to correct the orbital elements.
An illustration of orbital perturbations by the Earth's gravitational field is given in Fig.~\ref{fig:close-perturb}.
There is no significant perturbation to the orbits of impactors on a head-on or rear-end collision trajectory (models A1, C1, A3, and C3).
However, assuming a $\delta=90^\circ$ or a $\delta=270^\circ$ impact angle, the impactor's trajectory is strongly perturbed, see the departure of fly path in Fig.~\ref{fig:close-perturb} for models A2, C2, A4, and C4.

It should be noted that the A1 and C1 models are special configurations and that in both cases the impactor is in its apocentre position at the impact event.
In model A1, the Earth's orbital speed is higher than that of the impactor, so the Earth will run into the impactor.
In model C1, the impactor is travelling in a retrograde orbit, but the Earth's orbital speed is still higher than that of the impactor when impacting.

Orbital elements determined at one year before impact are given in Table~\ref{tab:Realorbelem}.
We emphasise that asteroid-type and one comet-type impactors resemble the Aten and Apollo group of NEOs which are indicated in Table~\ref{tab:Realorbelem}.





\begin{table}[H]
        \centering
        \caption{Orbital elements of asteroid-type (A1-A4) and comet-type (C1-C4) impactors determined at the impact event.}
        \label{tab:orbelem}
        \begin{tabular}{lcccccc} 
                model & $\delta\,(^\circ)$  & $\phi\,(^\circ)$ & $a$\,(au) & $q$\,(au) & Q\,(au) &  $e$ \\
\hline

        A1 & 0   & 0  & 0.534 & 0.053 & 1.015 & 0.900 \\
        A2 & 90  & 0  & 1.822 & 0.596 & 3.048 & 0.673 \\
        A3 & 180 & 0  & -1.295     & 1.016     & -     & 1.785 \\
        A4 & 270 & 0  & 1.821 & 0.603 & 3.039 & 0.669 \\
\hline
        C1 & 0   & 0  & 0.544   & 0.072 & 1.017 & 0.869 \\   
        C2 & 90  & 0  & -1.244  & 0.423     & -     & 1.340         \\
        C3 & 180 & 0  & -0.290  & 1.016     & -     & 4.502         \\
        C4 & 270 & 0  & -1.245  & 0.430     & -     & 1.345         \\
\hline
\end{tabular}

\end{table}

\begin{table}[H]
        \centering
        \caption{Orbital elements of asteroid-type (A1-A4) and comet-type (C1-C4) impactors determined at the interception date.}
        \label{tab:Realorbelem}
        \begin{tabular}{lcccccc} 
                model & $\delta\,(^\circ)$  & $\phi\,(^\circ)$ & $a$\,(au) & $q$\,(au) & Q\,(au) &  $e$ \\
\hline

        A1$^{\mathrm{(a)}}$ & 0   & 0  & 0.560 & 0.103 & 1.016 & 0.816 \\
        A2$^{\mathrm{(b)}}$ & 90  & 0  & 1.454 & 0.638 & 2.270 & 0.561 \\
        A3 & 180 & 0  & -2.433     & 1.016     & -     & 1.418 \\
        A4$^{\mathrm{(b)}}$ & 270 & 0  & 1.454 & 0.644 & 2.263 & 0.557 \\  
\hline
        C1$^{\mathrm{(a)}}$ & 0   & 0  & 0.534 & 0.051 & 1.016 & 0.904 \\   
        C2 & 90  & 0  & -1.502     & 0.437     & -     & 1.291         \\
        C3 & 180 & 0  & -0.312       & 1.017     & -     & 4.255         \\
        C4 & 270 & 0  & -1505      & 0.444     & -     & 1.295         \\
\hline
\\
\multicolumn{7}{l}{$^\mathrm{(a),(b)}$ Resembles the Aten and Apollo group of NEOs}
\end{tabular}
\end{table}

\begin{figure*}
    \centering
    \includegraphics[width=1.85\columnwidth]{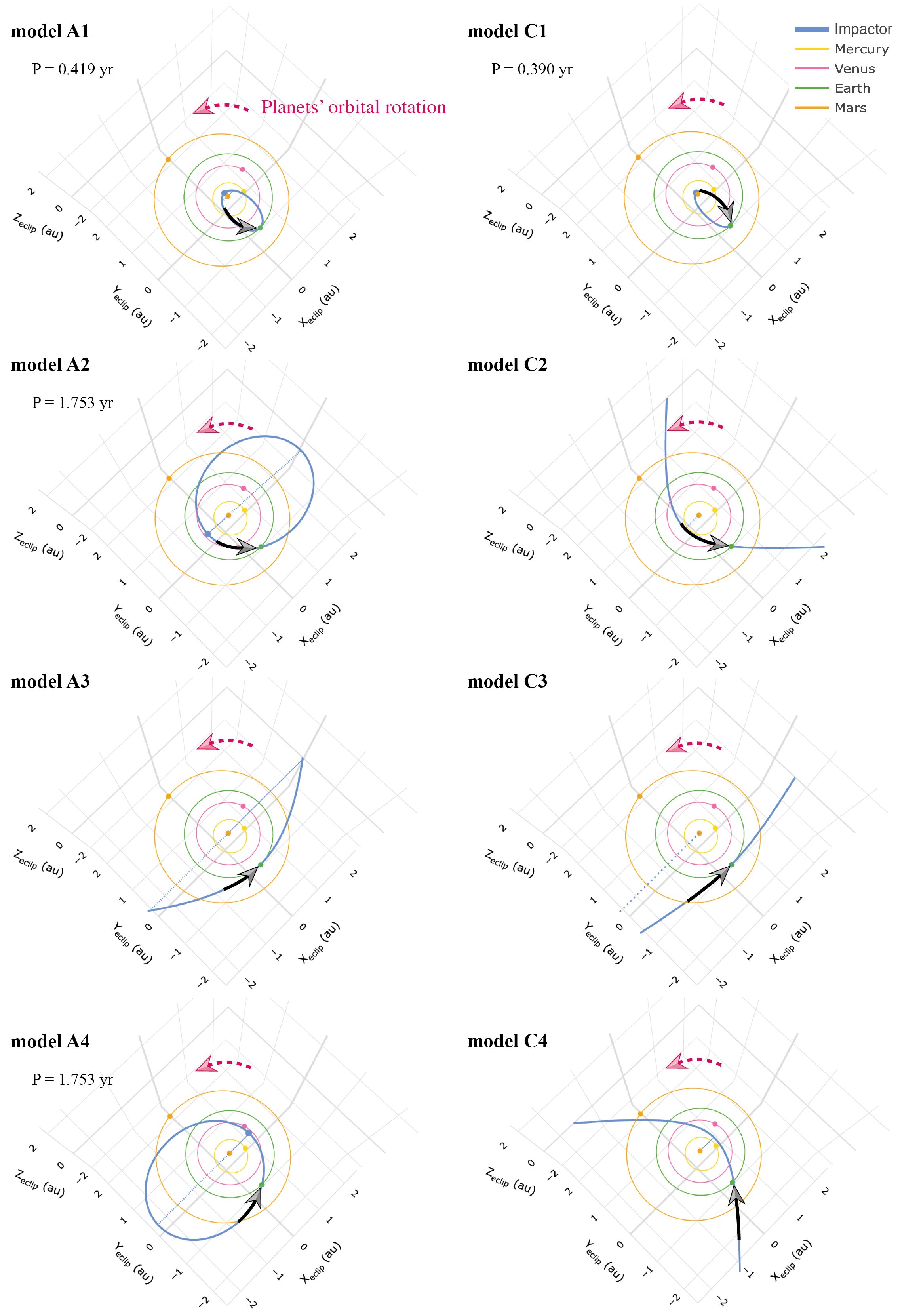}
    \caption{Visualisation of the orbits of the impactors (yellow): left for the asteroid-type, right for the comet-type impactors. The orbits of the inner planets are also shown. Red 
    arrow indicates the rotation of the planets around the Sun, while black 
    arrow indicates the fly path of the impactor.
    }
    \label{fig:orbit}
\end{figure*}

Figure~\ref{fig:orbit} shows the visualisation of the orbits of the hypothetical impactors whose orbital elements are presented in Table~\ref{tab:Realorbelem}. The orbits of planets of the solar system are also shown. The JPL Custom Orbital Visualization tool\footnote{\url{https://ssd.jpl.nasa.gov/tools/orbit_diagram.html}} was used to create the orbital plots.

\section{Integrator order and precision} 
\label{sec:apx-integrator}


The adaptive timestep, $dt^{(1)}$, is recalculated at each step according to \cite{NitadoriMakino2008} as
\begin{equation}
dt^{(1)}=\eta dt^{(0)} \mathrm{min}\left\{\left(\frac{a^\mathrm{pred}_i}{a^{(0)}_i}\right)^{1/p}\right\},
\end{equation}
where $dt^{(0)}$ is the timestep at the previous step, $a^\mathrm{pred}_i$ and $a^\mathrm{(0)}_i$ are the acceleration of the i-th body at the predictor and previous stage, and $p$ is the order of the integrator.

We conducted a series of integrator precision measurements to determine the optimal order with acceptable position error.
The impactor is modelled as a single body omitting the particle distribution generation.
All major bodies in the Solar System are included in the integration.
After 3\,years of backward and forward integration, the impactor should be at the same position as it initially was, that is, at the impact point.
The hypothetical impact event for the precision error measurements assumes an asteroid model with parameters of $\delta=0$, $\phi=0$, $v_\mathrm{imp}=20\,\mathrm{km\,s^{-1}}$.
Due to the finite number representation used for computation (32 bit double precision) and the applied order of the Hermite scheme, the calculation resulted measurable position error.  

\begin{figure}[H]
        \includegraphics[width=1.0\columnwidth]{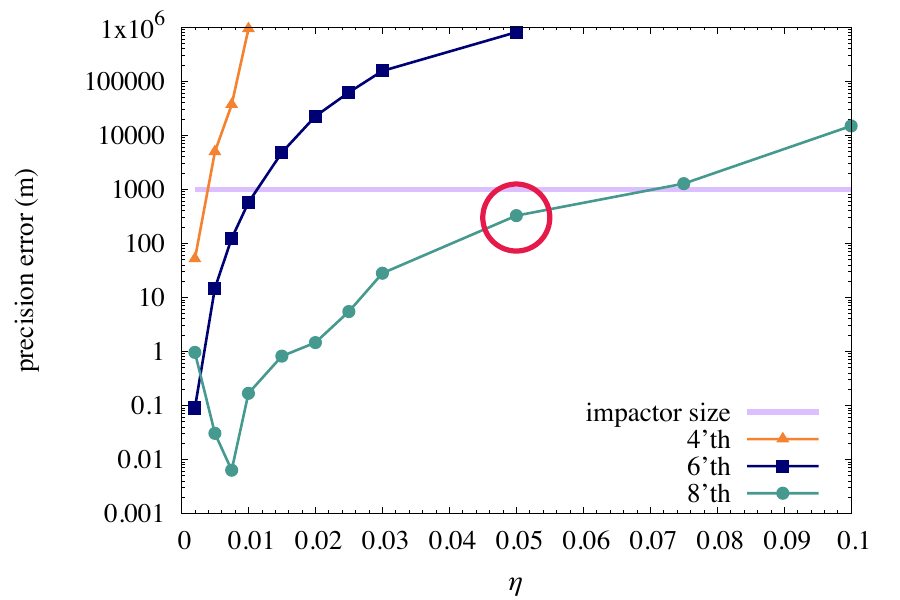}
    \caption{Measurements of position errors with fourth-, sixth-, and eighth-order Hermite schemes using different $\eta$ parameters for the adaptive timestep calculation. 
    The size of the impactor is also shown with a horizontal line.
    The optimal scheme with the desired precision (eighth order with $\eta=0.05$) is indicated with red circle.}
    \label{fig:prec}
\end{figure}

By measuring this position error, we can determine the precision of the applied integrator.
Figure\,\ref{fig:prec} shows our measurements for three different integrator orders.
With the smallest $\eta=0.001$ setting, the fourth-order scheme results in about 100 metre precision, while the sixth and eighth order gives about a metre precision. 
We note, however, that $\eta=0.001$ results in a very small time steps, that is, extremely slow integration.
With $\eta=0.05$ the eighth-order scheme gives a position error below the size of the impactor (1000 m).
However, using the fourth- or sixth-order scheme the precision grows several orders above the size of the impactor.
Thus, to integrate with a relative position error less then the physical size of the impactor, the eighth-order scheme is selected with $\eta=0.05$ throughout this study.

\section{Size and shape of the fragment cloud}
\label{apx:triax}

\begin{figure}[H]
    \centering    
    \includegraphics[width=0.9\columnwidth]{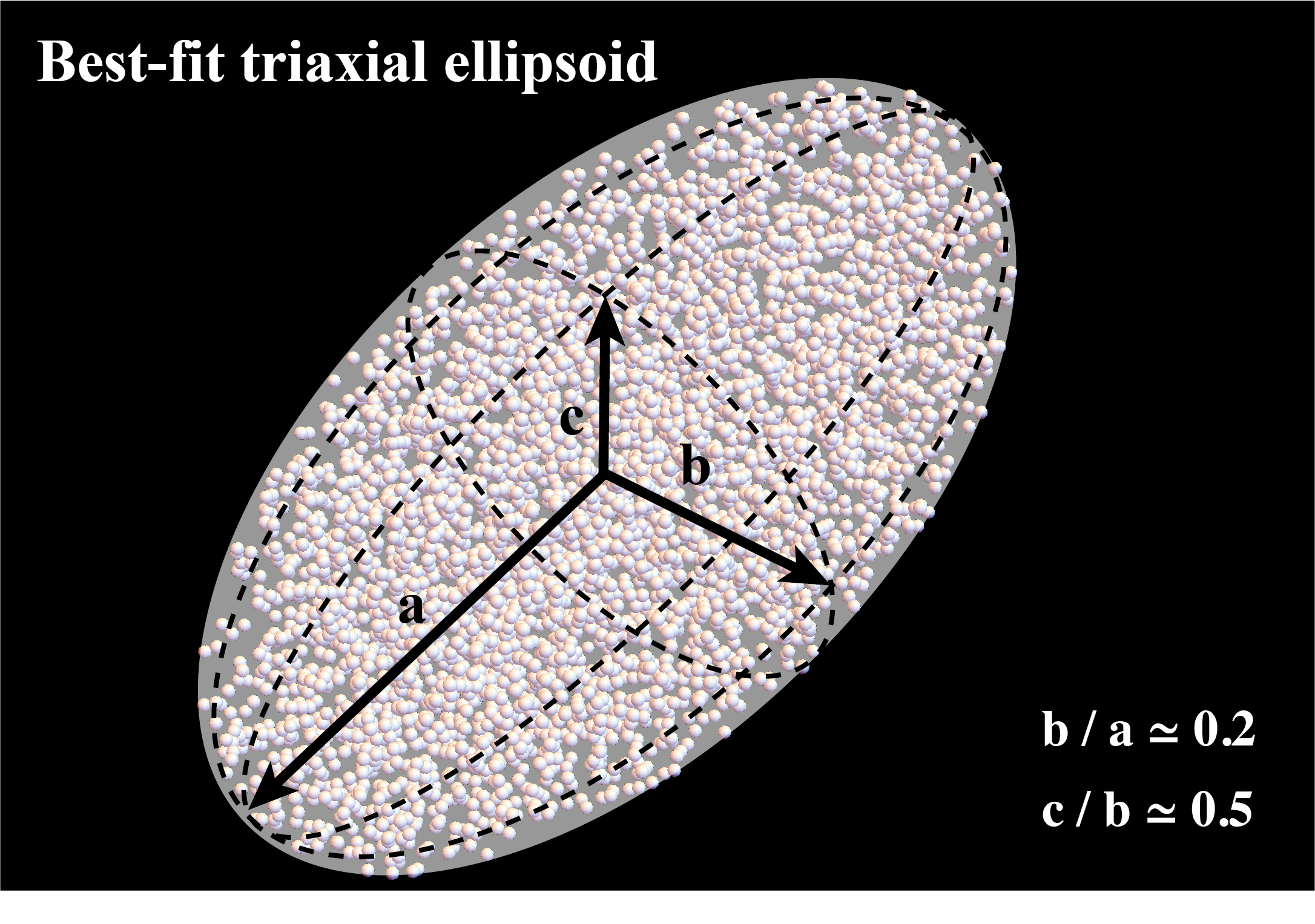}
    \caption{Example of the best-fit triaxial ellipsoid model of a fragment cloud: $a$ is the semi-major axis, and $b,\,c$ are the two semi-minor axes of the ellipsoid. The fragment distribution is taken from model C4.}
    \label{fig:best-fitell}
\end{figure}

To describe the shape of the fragment cloud, a best fit ellipsoid is determined for its particles' distribution (see an example in Fig~\ref{fig:best-fitell}).
This is done by calculating the cloud's moment of inertia tensor.
The routine {\tt def-triax.c} is based on the well-known Jacobi eigenvalue algorithm\footnote{\url{https://en.wikipedia.org/wiki/Jacobi_eigenvalue_algorithm}}, which is an iterative method of rotation for the calculation of the eigenvalues and eigenvectors of a real symmetric matrix. 
For a review, we refer to the paper by \cite{GV2000} on the topic of efficient and high performance eigenvalue  computation algorithms.

For each snapshot, we determined the half mass radius r$_{\rm hm}$ of the particle distribution (simply sorting by the individual distances of the particles in the cloud from the centre of the mass of the whole system). In our further shape investigation, we used only the particles which we find inside the $4\times r_{\rm hm}$. 

In our case, the matrix of inertia (MOI) in a centre–of–mass frame for the set of N particles by definition is a symmetric and real matrix: 
\begin{equation}
{\rm I}_{\rm com} \equiv
\left[
\begin{array}{ccc}
 \sum_{\mathrm{i}} m_\mathrm{i} (y_\mathrm{i}^2 + z_\mathrm{i}^2) & -\sum_{\mathrm{i}} m_\mathrm{i} x_\mathrm{i}^2 y_\mathrm{i}^2     & -\sum_{\mathrm{i}} m_\mathrm{i} x_\mathrm{i}^2 z_\mathrm{i}^2 \\
-\sum_{\mathrm{i}} m_\mathrm{i} y_\mathrm{i}^2 x_\mathrm{i}^2     &  \sum_{\mathrm{i}} m_\mathrm{i} (x_\mathrm{i}^2 + z_\mathrm{i}^2) & -\sum_{\mathrm{i}} m_\mathrm{i} y_\mathrm{i}^2 z_\mathrm{i}^2 \\
-\sum_{\mathrm{i}} m_\mathrm{i} z_\mathrm{i}^2 x_\mathrm{i}^2     & -\sum_{\mathrm{i}} m_\mathrm{i} z_\mathrm{i}^2 y_\mathrm{i}^2     &  \sum_{\mathrm{i}} m_\mathrm{i} (x_\mathrm{i}^2 + y_\mathrm{i}^2) \\
\end{array}
\right]
.\end{equation}
Here $m_\mathrm{i}$, $x_\mathrm{i}$, $y_\mathrm{i}$, and $z_\mathrm{i}$ stand for the $i$-th particle's mass and Cartesian coordinates.
A summation is performed for all particles found inside $4\times r_{\rm hm}$

After the Jacobi rotations (which in the end define for us a resulting first Euler angle $\theta$), the MOI matrix contains only diagonal components:
\begin{equation}
{\rm I}^{\alpha}_{\rm com} =
\left[
\begin{array}{ccc}
 {\rm I}_{xx} &       0       &       0 \\
      0       &  {\rm I}_{yy} &       0 \\
      0       &       0       &  {\rm I}_{zz} \\
\end{array}
\right]
\end{equation}

In the next step, we find the maximum - ${\rm I}_{max}$, middle - ${\rm I}_{mid}$ and minimum - ${\rm I}_{min}$ values from these diagonal elements. 
The axis ratio of the particle distribution ellipsoid can then easily be defined as: 
\begin{equation}
b/a = \sqrt{ {\rm I}_{mid} / {\rm I}_{max} }
\;\;\;\;\;\;\;
c/a = \sqrt{ {\rm I}_{min} / {\rm I}_{max} }
\end{equation}

The actual code is available from the authors upon an email request.

\section{Impact number, analytical estimation}
\label{apx:analytic}

\begin{figure}[H]
    \centering
    \includegraphics[width=0.9\columnwidth]{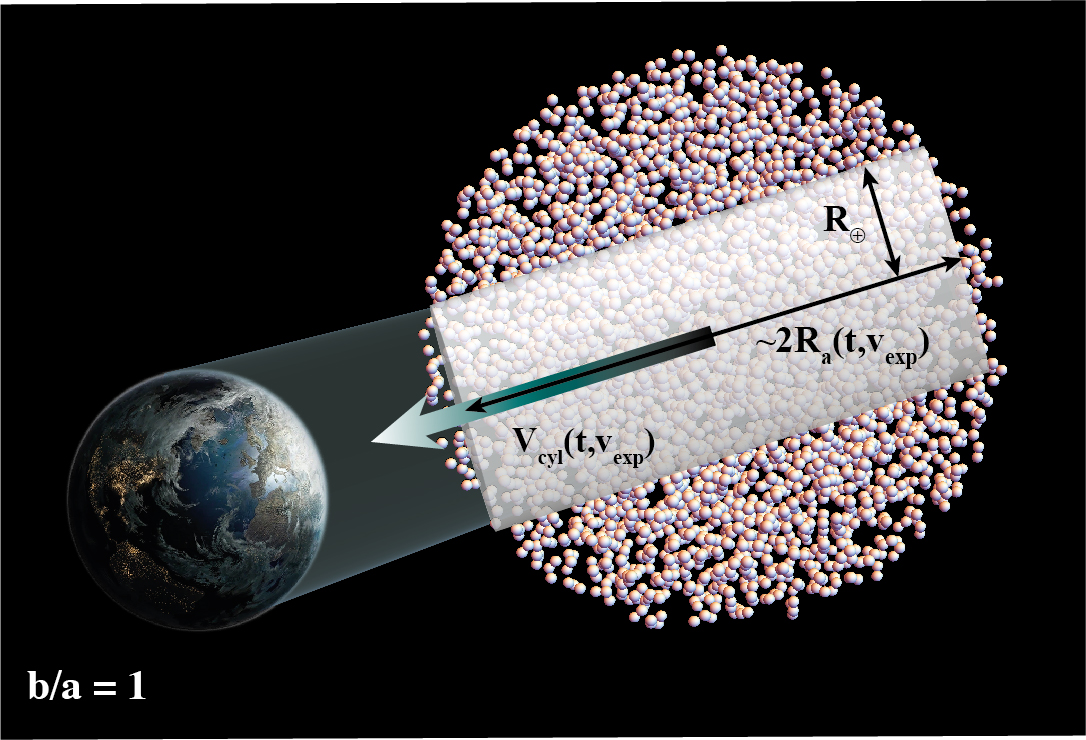}
    \caption{Fragment cloud before impact. The number of impacts can be estimated by counting the number of fragments that are inside the white cylinder.}
    \label{fig:analytic_model}
\end{figure}

The number of fragments that hit Earth, $N_\mathrm{imp}$, can be estimated by counting the fragments in a spherically symmetric homogeneous fragment cloud model, see Fig.~\ref{fig:analytic_model}.
Here, we re-introduce the method presented in \citet{Lubinetal2023}.
For simplicity, orbital dynamics (which can distort the spherical symmetry) and self-gravity (which can distort the homogeneity) are neglected in this model.
The radius of the expanding cloud of fragments is given by $R_\mathrm{clo}(t, v_0) = R_\mathrm{0}+v(D/2)t$ at a given time $t$, where $v(D/2)$ is the speed at which the cloud expands at the surface, and an $D$ is the initial diameter of the impactor.
Assuming a completely spherical fragment cloud, its volume will be $V_\mathrm{clo}(t,v_0)=(4/3) \pi R_\mathrm{clo}(t, v_0)^3$ at a given time.
The cylindrical section (whose diameter is equal to that of the Earth, represented by the white area in Fig.~\ref{fig:analytic_model}) is $V_\mathrm{cyl}(t, v_0) = R_\oplus^2 \pi 2 R_\mathrm{clo}(t, v_0)$, where $R_\oplus$ is Earth's radius.
Assuming a homogeneous distribution of fragments in the cloud, the impact number (i.e. the number of of fragments in the cylinder) is:
\begin{equation}
N_\mathrm{imp} \simeq N_\mathrm{tot}\frac{V_\mathrm{cyl}(t, v_0)}{V_\mathrm{clo}(t,v_0)}=N_\mathrm{tot}\frac{3}{2}\frac{R_\oplus^2}{R_\mathrm{clo}(t, v_0
)^2},
\label{eq:Nimp}
\end{equation}
with $N_\mathrm{tot}$ being the total number of fragments in the cloud.
We note that due to the non–uniform expansion, the assumption of a homogeneous distribution of the fragment cloud fails for gravitationally non-self-interacting models with $v_0=1$, but can be plausible for gravitationally interacting cloud models. 

\section{Simulations on interception dates} 
\label{sec:apx-sim}

\begin{figure*}
        \includegraphics[width=2\columnwidth]{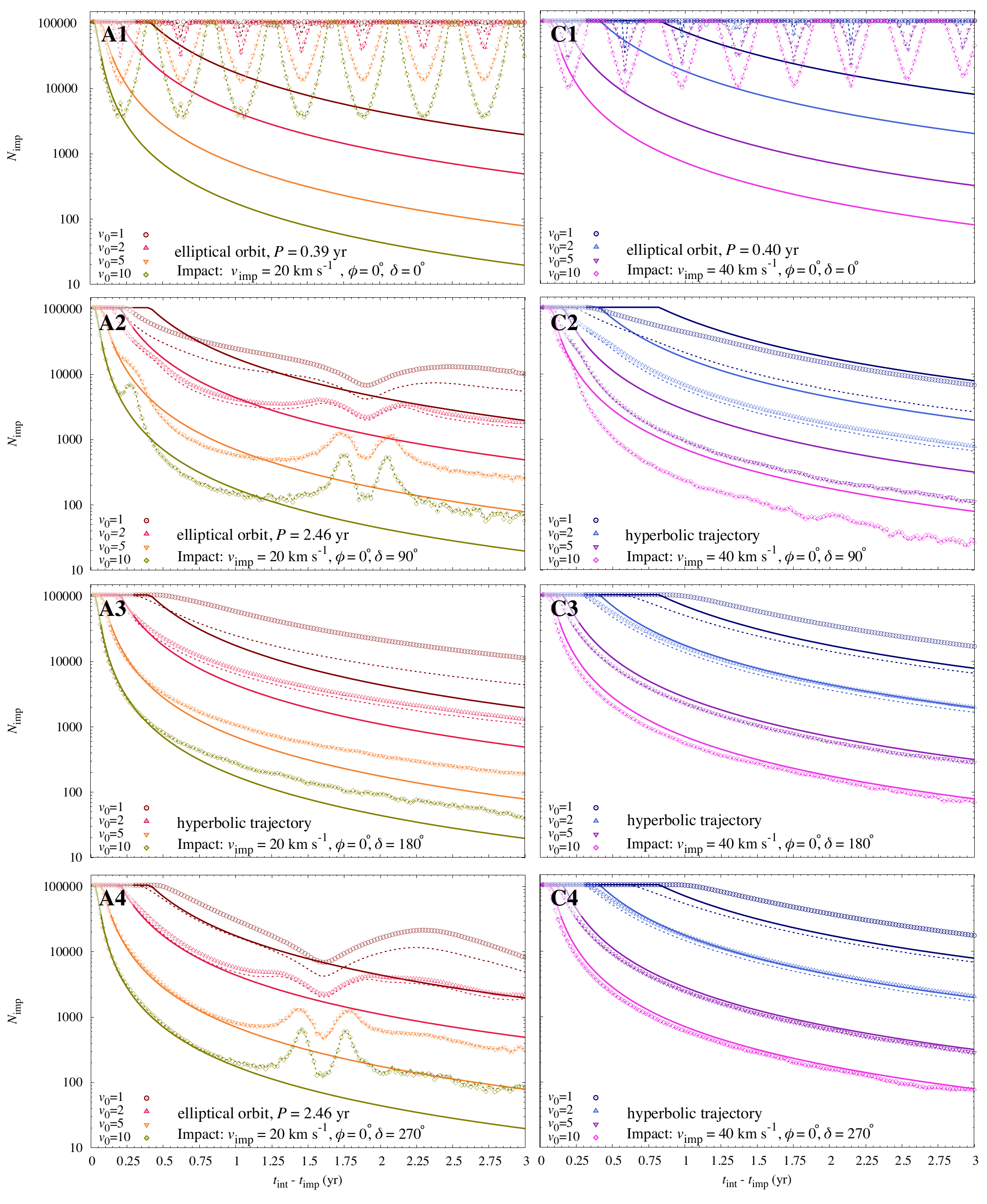}
    \caption{Number of fragments that hit Earth, $N_\mathrm{imp}$, as a function of $t_\mathrm{imp}-t_\mathrm{int}$ for a set of impact models (assuming longitude of $\delta=0^\circ,\,90^\circ,\,180^\circ$ and $270^\circ$, see model details in Table\,\ref{tab:param}).
    Left and right panels show asteroid-type (models A1-A4) and comet-type (models C1-C4) impactors, respectively.
    Models assume four cloud expansion velocities, $v_0$ indicated with four distinct colours. 
    The warm and cool colours represent the asteroid-type ($v_\mathrm{imp}=20\,\mathrm{km\,s^{-1}}$) and comet-type ($v_\mathrm{imp}=40\,\mathrm{km\,s^{-1}}$) impactors. The gravitationally interacting models are shown with symbols. 
    The non-interacting impact cloud models are shown with thin solid and thin dashed lines, respectively. 
    Thick solid lines represent the theoretical impact number given by equation\,\ref{eq:Nimp}, for which case the effect of orbital dynamics on the cloud and the fragment's self-gravity are neglected.}
    \label{fig:Nimp}
\end{figure*}

The number of impacts on Earth as a function of $\Delta t = t_\mathrm{int}-t_\mathrm{imp}$ with different values of
expansion velocity, $v_0=1,\,2,\,5$, and $10$ is presented in Fig.~\ref{fig:Nimp}. 
All asteroid-type (A1-A4) and comet-type (C1-C4) models are computed assuming impact direction $\delta=0^\circ,\,90^\circ,\,180^\circ$, and $270^\circ$. Self-gravitating and non-self-gravitating models are also compared with the analytical predictions. 
Discussion is given in Sect.~\ref{sec:int-date}.

\section{Simulations on rotation axis} 

\begin{figure*}
    \center
        \includegraphics[width=1.9\columnwidth]{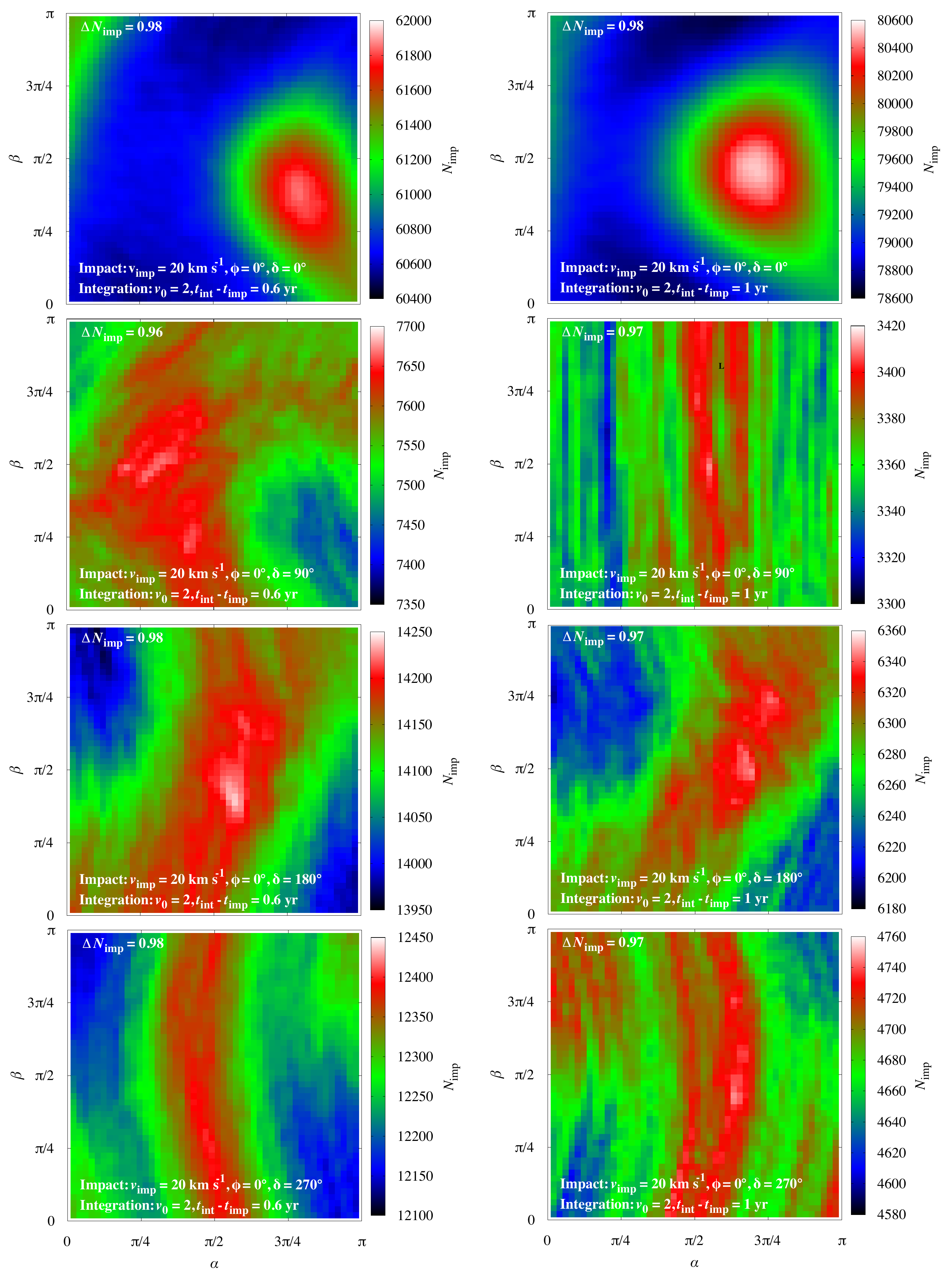}
    \caption{Effect of the orientation of the rotational axis on the number of fragments that hit Earth, $N_\mathrm{imp}$. The minimum rotational period is assumed (2.04 hours for a rubble pile composition corresponding to $0.42\,\mathrm{m\,s^{-1}}$ surface velocity) for an asteroid–type impactor. The integration length is 0.6 and 1 years for left and right panels, respectively. Models A1-A4 (impact angles of $\phi=0$ and $\delta=0^\circ\,,90^\circ,\,180^\circ,$ and $270^\circ$) are investigated. The ratio of the minimum and maximum impact numbers, $\Delta N_\mathrm{imp}$, are calculated for each model.}
    \label{fig:Ast-spin}
\end{figure*}

\begin{figure*}
    \center
        \includegraphics[width=1.9\columnwidth]{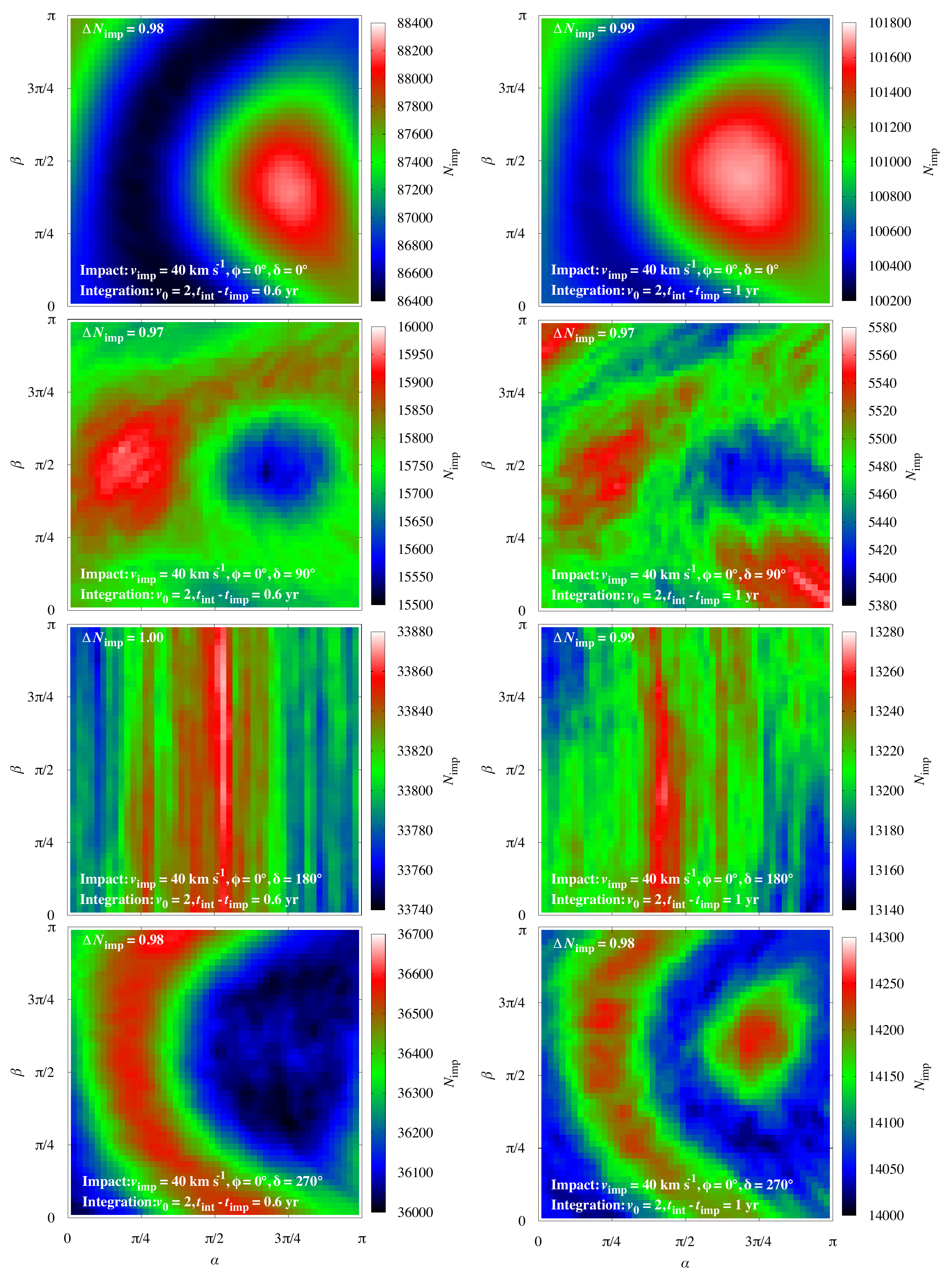}
    \caption{Same as Fig.~\ref{fig:Ast-spin} for a comet-type impactor, i.e. computations for models C1-C4. In this case the minimum rotational period is 4.1 hours and the corresponding surface velocity is $0.21\,\mathrm{m\,s^{-1}}$.}
    \label{fig:Comet-spin}
\end{figure*}

Additional runs of non-self-gravitating fragment cloud
models assuming $v_0=2$ which take into account the rotation of the impactor are presented in Figs~\ref{fig:Ast-spin} and \ref{fig:Comet-spin}. 
A discussion is given in Sect.~\ref{sec:rot}.

\label{lastpage}
\end{document}